\documentclass[11pt]{article}

\usepackage[preprint]{acl}

\usepackage{times}
\usepackage{latexsym}

\usepackage[T1]{fontenc}

\usepackage[utf8]{inputenc}

\usepackage{microtype}

\usepackage{inconsolata}

\usepackage{graphicx}
\usepackage{xcolor}
\usepackage{subcaption}

\usepackage{amssymb}
\usepackage{amsmath}
\usepackage{mathtools}

\usepackage{enumitem}
\setlist{nolistsep}

\usepackage{algorithm}
\usepackage{algorithmic}

\usepackage{url}
\usepackage{hyperref}

\usepackage{booktabs}
\usepackage{multirow}

\newcommand{\secref}[1]{\S\ref{#1}}

\title{HALvest-Contrastive: Retrieval-Like Authorship Attribution \\with Patch-level Late Interaction} 

\author{Francis Kulumba \\
  Inria Paris \\
  Sorbonne Université \\
  \texttt{francis.kulumba@inria.fr} \\\And
  Wissam Antoun \\
  Inria Paris \\
  Sorbonne Université \\\AND
  Guillaume Vimont \\
  IRIF \\\And
  Laurent Romary \\
  Inria Paris \\\And
  Florian Cafiero \\
  LRE, EPITA \\
  Ecole nationale des chartes -- PSL}

\begin{document}

\maketitle

\begin{abstract}
    Authorship attribution asks whether two pieces of text share a writer, but topical confound makes the task deceptively easy: two authors covering the same topic may look more alike than one author covering two topics. Scholarly prose offers a natural remedy, academic writers produce multiple papers on related but distinct topics while maintaining consistent stylistic habits. We introduce HALvest, a 17-billion-token multilingual corpus of open-access academic papers, and its English contrastive derivative HALvest-Contrastive, where same-author passages are drawn from distinct papers within a disciplinary field to minimize topical overlap. We validate our benchmark by showing that a strong lexical baseline collapses once topical shortcuts are removed. On this same benchmark, we revisit how authorship is scored. Standard systems compress each document into a single vector. We instead keep a sequence of vectors and compare them with late interaction, then propose patch-level late interaction, which groups neighboring tokens into patches before matching. Matching at the sequence level greatly improves performance over the single-vector baseline, but the optimal interaction granularity is subtle.
\end{abstract}

\section{Introduction}
\label{sec:introduction}

\begin{figure*}[t]
\centering
\includegraphics[width=\textwidth]{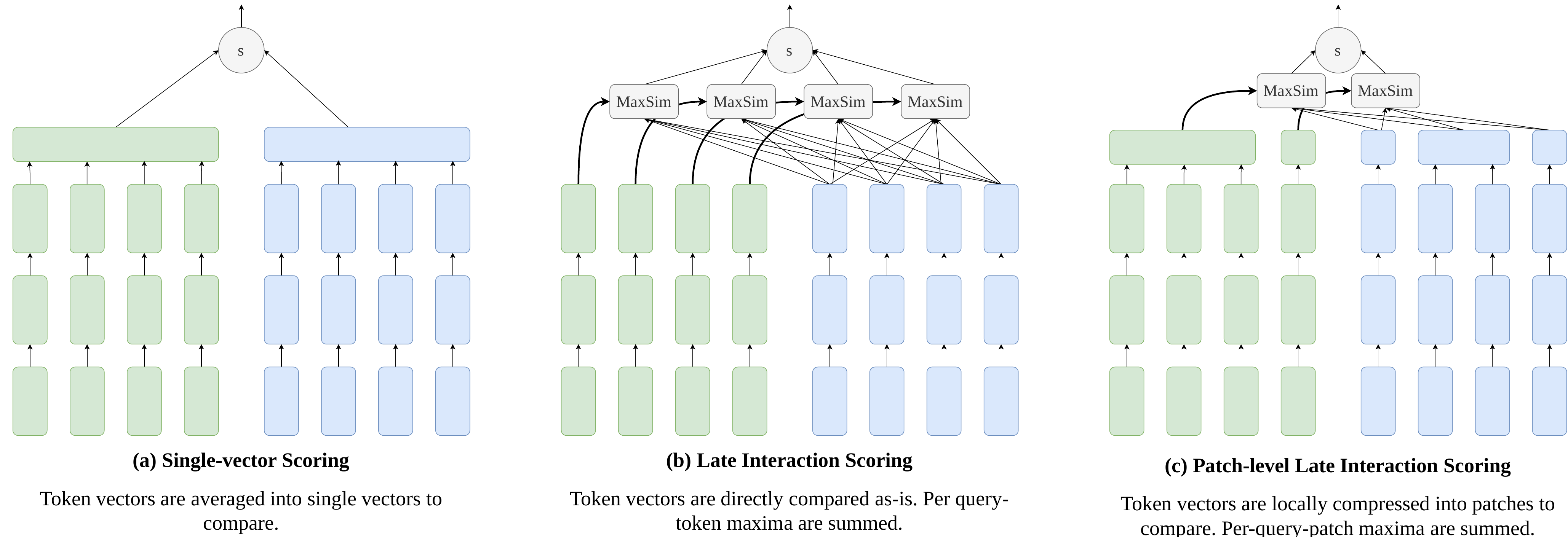}
\caption{\textbf{Three scoring modes for authorship attribution.} The green and blue tiles represent token vectors from an anchor and a positive document respectively. \textbf{(a) Single-vector pooling} averages token embeddings into one vector per document and scores pairs by cosine similarity. \textbf{(b) Late interaction}~\citep{khattab_colbert_2020} keeps the full sequence of token vectors and scores by $\mathrm{MaxSim}$. \textbf{(c) Patch-level late interaction (PLI)} groups token vectors into contiguous patches, pools each patch into a single vector, and applies $\mathrm{MaxSim}$ at the patch level. We train and evaluate all three modes under a matched contrastive objective. The granularity comparison is the main object of study in this paper.}
\label{fig:scoring_modes}
\end{figure*}
 
Authorship Attribution (AA) asks whether two pieces of text were written by the same author. The field has a long statistical tradition that treats style as a document-level property summarized by a frequency vector~\citep{burrows_delta_2002, schler_effects_2006, treeratpituk_disambiguating_2009}. Neural approaches have largely inherited this view. A passage is embedded by a transformer encoder~\citep{vaswani_attention_2017, devlin_bert_2019}, the token representations are mean-pooled into a single vector, and training minimizes a triplet loss: a three-way comparison between an anchor, a same-author positive, and a different-author negative, with a fixed margin~\citep{wegmann_same_2022, kantharuban_idiolex_2026}. Information retrieval, a field that also contrasts documents against one another, has moved in a slightly different direction. In fact, dense retrieval has replaced triplet loss with InfoNCE~\citep{oord_representation_2019, gao_simcse_2021}, a contrastive objective that leverages many in-batch negatives instead of one hard negative, and single-vector pooling has given way to late interaction (LI)~\citep{khattab_colbert_2020, santhanam_colbertv2_2022}, where documents are represented by sets of token vectors and the score is the sum of maximum matches between said tokens (Figure~\ref{fig:scoring_modes}). This paper asks what happens when the retrieval playbook is applied to AA, and where the two tasks diverge.
 
A primary obstacle is topical confounding: if the AA field moved toward contrastive approaches, it was, to a large extent, to be able to minimize vocabulary overlap within each triplet~\citep{wegmann_does_2021}. The intuition is that decreasing the discriminative signal from individual tokens may encourage models to learn subtler distributional cues like stop-word frequency, sequence length, or overall tone. Building on that same line of work, we propose a solution based on the inherent nature of academic papers, where authors write on related but distinct scholarly topics while maintaining consistent stylistic patterns. We introduce HALvest, a 17-billion-token corpus harvested from 778k open-access and multilingual academic papers, and its contrastive derivation HALvest-Contrastive, an English-only benchmark with controlled topic variation. We built HALvest-Contrastive to maximize topic decoupling: an anchor and positive are drawn from \emph{different} papers written by the same author-set, and the negative is mined from within the same disciplinary field, so that a model cannot entirely succeed by learning topic vocabulary.
 
Under this design, we study how retrieval techniques transfer to authorship attribution, from the training objective to the interaction mechanism (late interaction and its patch-level variant). The comparison is controlled: all models share the same encoder, the same data, and the same loss; only the pooling and scoring granularity differ. Our contributions are as follows.
 
\begin{enumerate}
    \item \textbf{We release HALvest}, a 17-billion-token multilingual corpus harvested from open-access academic papers, and \textbf{HALvest-Contrastive}, an English scholarly AA benchmark with verified multi-author structure, controlled topic decoupling, and an evaluation structure by span length.
    \item \textbf{We empirically validate topic decoupling} by showing that a strong lexical baseline succeeds on topic-rich but fails on topic-decoupled data, with a gap persisting across span lengths.
    \item \textbf{We present a granularity study} that pits mean pooling, token-level late interaction, and fixed-policy PLI variants against one another on matched data under a matched training objective, across five span-length subsets of HALvest-Contrastive plus PAN19 zero-shot.
    \item \textbf{We observe that the empirically best fixed-$n$ patch size} across our subsets and PAN19 is well approximated by $n_\star \approx 0.18 \cdot \sqrt{S}$, where $S$ is the token count. A descriptive fit that provides a practical starting point for practitioners.
\end{enumerate}
 
The code~\footnote{\url{https://github.com/Madjakul/DeepStylometry}}, and the corpus itself~\footnote{\url{https://huggingface.co/datasets/almanach/halvest-contrastive}} are released under an open license.

\section{Related Work}
\label{sec:related}

The closest precedents to our setup come from three separate literatures: stylometry, scholarly text corpora, and multi-vector retrieval.

\subsection{Stylometry and authorship attribution.}

Stylometric AA has traditionally relied on frequency-based representations of writing style, leveraging features such as function words~\citep{mosteller_inference_1963, burrows_delta_2002, treeratpituk_disambiguating_2009}, part-of-speech n-grams~\citep{argamon_style_1998}, most frequent character n-grams~\citep{keselj_ngram_2003}, or combinations of those~\citep{cafiero_why_2019, juola2015rowling}. Machine-learning treatments formalised the task in closed, verification, and open-set variants~\citep{koppel_computational_2009, seroussi_authorship_2014}. A recurring concern is topical confounding: \citet{sapkota_not_2015} showed that classifiers trained and tested on the same topic substantially overestimate generalization, and proposed cross-topic evaluation as a stricter protocol.
 
Large-scale neural AA systems followed: contrastive learning has emerged as a powerful paradigm for stylometry~\citep{ai_whodunit_2022, wegmann_same_2022, huertas-tato_isolating_2024}. A prevailing strategy in this area has been to decorrelate an author's style from literal text content~\citep{altakrori_topic_2021}. This includes developing edit-invariant loss functions to capture signals beyond surface-level text~\citep{liu_dont_2022}, or creating positive pairs with high semantic overlap to force models to learn distributional cues~\citep{rivera-soto_learning_2021, wegmann_does_2021, alshomary_layered_2025}. Most recently, \citet{kantharuban_idiolex_2026} studied idiolect at scale.

Our work differs in two ways. First, it targets scholarly prose, a low-entropy register with strong domain structure, rather than internet or literary text. Second, it replaces single-vector pooling and triplet loss with late interaction and InfoNCE~\citep{oord_representation_2019}, and asks whether the induced performance changes transfer (Figure~\ref{fig:scoring_modes}). In fact, the InfoNCE objective has become a standard discriminative loss for representation learning: it learns embeddings by contrasting a positive pair against a pool of in-batch negatives. \citet{chen_simple_2020} and \citet{Khosla_supervised_2021} showed that large negative pools and strong augmentations suffice for self-supervised image representations. Notably, for our setting, they report that in-batch false negatives, pairs labeled as negative that in fact share a latent class, degrade performance only weakly.
 
\subsection{Scholarly text corpora.}

Several corpora have been built around scientific writing. Prominent examples include domain-specific documents from ArXiv, DBLP, and PubMed~\citep{sen_collective_2008, dogan_ncbi_2014, wahle_d3_2022}. Other initiatives provide full-text access, such as the ACL Anthology for computational linguistics~\citep{bird_acl_2008}, or S2ORC~\citep{lo_s2orc_2020}, notable for its scale and multi-domain coverage.

HALvest complements these by drawing from a single open-access repository that spans multiple high-level domains, and by supplying the author-set labels as they appear in the metadata used by real retrieval systems.
 
\subsection{Late interaction and multi-vector retrieval.}

ColBERT~\citep{khattab_colbert_2020} introduced late interaction: documents are represented by a set of token embeddings and scored against a query via $\mathrm{MaxSim}$, the sum over query tokens of the maximum cosine similarity to any document token. ColBERTv2~\citep{santhanam_colbertv2_2022} added residual compression for storage. TRIAL~\citep{kang_trial_2025} adds bigram-level token relation scores and per-query importance weights to $\mathrm{MaxSim}$. These modifications target semantic retrieval where phrase coherence and content-word emphasis improve relevance estimation. Token pruning approaches~\citep{zong_towards_2025} learn to remove tokens before scoring. ColBERTer~\citep{hofstatter_introducing_2022} compresses tokens into coarser units, while ConstBERT~\citep{macavaney_efficient_2025} pools token embeddings into a fixed-size set of learned vectors, functioning as a learned-patch reranker. 

These methods are conceptually the closest precedents to our patch-level view, but they target either semantic retrieval or generative modeling rather than AA.

\section{HALvest and HALvest-Contrastive}
\label{sec:dataset}

A benchmark for stylometry has to answer two questions: where does the text come from, and how are same-author and different-author pairs constructed so that the resulting signal is style rather than topic. This section answers both.

\subsection{Why scholarly prose?}

Scholarly writing is an under-exploited testbed for stylometric research. In fact, academic prose is low-entropy: its vocabulary is rather constrained, its syntactic templates are strongly standardized, and its structure is recurrent across papers within a field. This compresses the range of surface variation and makes the stylistic residual easier to isolate from topical noise.
 
\subsection{Source and processing pipeline.}

HAL is an open-access repository of French and international scholarly output. We extract full-text PDFs for all papers with permissive licenses, process them with GROBID~\citep{GROBID} to recover structured XML, which we serialize to plain text. All in all, HALvest covers 778k documents in 56 languages across 16 disciplinary domains.

\paragraph{From HALvest to HALvest-Contrastive.}
For the contrastive dataset we apply an additional sequence-level filter that restricts to English spans; this yields HALvest-Contrastive, the English-only triplet dataset on which all models in this paper are trained. We further filter to the 13 domains with sufficient document counts for reliable hard-negative mining (Appendix~\ref{app:hal-stats}), and remove spans with excessive symbols, abnormal layout, code fragments or HTML markup, high uppercase ratios, or anomalous sentence-length variance. Mining proceeds in three configurations at first, before scaling the best one to millions of rows.
 
The \textbf{unrestricted} configuration samples five random spans per query document. Positives are drawn from all documents sharing the exact same author-set; if no other document exists for that set, positive spans are sampled from the same document as the query. Hard negatives are drawn from documents that share no authors with the query but have the same HAL domain label. This configuration permits high topical overlap between query and positive, and serves as a topical control.
 
The \textbf{base} (restricted) configuration enforces strict topic decoupling: positives are drawn \emph{exclusively} from different documents by the same author-set, never from the query document itself. The base configuration is the one we scaled and released under the name HALvest-Contrastive. The total number of triplets is described in Table~\ref{tab:contrastive_composition}.

\begin{table}[ht]
    \centering
    \begin{tabular}{cccc}
        \toprule
        \textbf{\#~Sentences}   & \textbf{Train} & \textbf{Valid} & \textbf{Test} \\
        \midrule
        2 & 1.88M & 19.1k & 19.1k \\
        4 & 1.4M & 14.3k & 14.3k \\
        6 & 1.11M & 11.3k & 11.3k \\
        8 & 892k & 9.1k & 9.1k \\
        \bottomrule
    \end{tabular}
    \caption{Number of triplets in the HALvest-Contrastive dataset.}
    \label{tab:contrastive_composition}
\end{table}
 
A third \textbf{inverse cloze task} (ICT) configuration~\citep{lee_latent_2019} treats a random span from a document as the anchor and the surrounding context as the positive, with a non-overlapping passage from the same document as the negative. ICT supplies a style-aware retrieval signal that is neither purely lexical nor fully topic-decoupled, and serves as an intermediate baseline in \secref{sec:exp-topic-decoupling}.

\paragraph{Triplet construction.}
Within each configuration, a triplet consists of three spans each containing $k$ contiguous sentences. The span length $k$ parametrizes task difficulty: shorter spans leave less textual evidence, so AA accuracy rises monotonically with $k$. The dataset supports $k \in \{2, 4, 6, 8, 10\}$; we report $k = 4$ as the primary split, with full per-$k$ results in Appendix~\ref{app:per-split}. Training uses base configuration triplets throughout; evaluation reports both base and unrestricted where relevant to the argument.

\paragraph{Author-set as the label.}
The label of a triplet is its set of co-authors, not a single ``primary'' writer and author. We use author-sets because they match the metadata exposed by scholarly repositories and avoid imposing an arbitrary single-author heuristic, thus being closer to real-world settings. This also covers the common case, since about 51\% of our triplets are already single-author, while preserving collaborative papers without forcing a noisy reduction to one writer. Forensic linguistics supports this framing. \citet{dauber_git_2019} demonstrate that individual attribution within collaborative documents achieves subpar accuracy even with best-available methods. In the general case we cannot know which listed co-authors contributed as writers versus providing ideas or feedback.

\paragraph{Cross-domain evaluation via PAN19.}
We use PAN19~\cite{kestemont_2019_3530313}, which labels documents at the individual-author level and whose texts are drawn from fan-fiction, as a zero-shot cross-domain evaluation. Both corpora are English, but the registers are different: academic prose is heavily normalized by journal style guides, while fan-fiction exhibits broader variation in sentence length, punctuation, and paragraph structure. A model that generalizes across both the label granularity gap (author-set to individual author) and the register gap (academic to fan-fiction) is, by elimination, not a topic classifier in disguise.

\section{Contrastive Modelling}
\label{sec:method}

We separate two design choices that the AA literature has historically bundled together. In this section, we make the case for InfoNCE over triplet loss as the training objective, before laying out the three families of pooling and interaction we compare in our experiments.

We adopt standard retrieval terminology throughout. A \emph{triplet} is an ordered tuple of three passages $(a, p, n)$, where $a$ (the \emph{anchor}) and $p$ (the \emph{positive}) share authorship, and $n$ (the \emph{negative}) does not. A contrastive loss pulls $a$ and $p$ close in embedding space while pushing $a$ away from $n$. We evaluate models with the standard ranking metrics Recall@$k$ (the fraction of anchors whose correct positive is in the top-$k$ retrieved candidates) and nDCG@$k$ (a graded, position-aware variant). For zero-shot cross-domain evaluation, we use the \emph{triplet accuracy} (the fraction of triplets for which $s(a, p) > s(a, n)$).

\subsection{A simple look on InfoNCE}
\label{sec:method-infonce}
 
Given an anchor $a$, a positive $p$ drawn from the same author-set, and a pool of in-batch negatives $\mathcal{N}$ drawn from other author-sets, we train to minimize the InfoNCE loss

\begin{equation}
\small
\mathcal{L} = -\log \frac{\exp(s(a, p) / \tau)}{\exp(s(a, p) / \tau) + \sum_{n \in \mathcal{N}} \exp(s(a, n) / \tau)}
\label{eq:infonce}
\end{equation}

where $s(\cdot, \cdot)$ is the pooling-specific similarity score (Figure~\ref{fig:scoring_modes}) cosine for single-vector pooling, $\mathrm{MaxSim}$ for late interaction and PLI, and $\tau$ is a temperature parameter, set to 0.5 throughout. We use full-gather across GPUs, so that the negative pool at each anchor grows with the total effective batch size. We used 4 GPUs with a batch-size of 32 per device, leading to an effective number of 256 negatives per anchor. We further scale the batch size to 64, as we show in \secref{sec:experiments}, this scaling is not incidental: larger negative pools yield measurable gains, consistent with the retrieval literature~\cite{karpukhin_dense_2020, chen_simple_2020}.
 
\paragraph{InfoNCE gradient naturally concentrates on hard negatives,} which aligns with suppressing topical shortcuts in authorship attribution.. Writing $s_n = s(a, n)$ for a generic negative, the partial derivative of Equation~\ref{eq:infonce} with respect to $s_n$ is

\begin{equation}
\footnotesize
\frac{\partial \mathcal{L}_{\mathrm{InfoNCE}}}{\partial s_n} \propto \frac{\exp(s_n / \tau)}{\exp(s_p / \tau) + \sum_{n' \in \mathcal{N}} \exp(s_{n'} / \tau)}
\label{eq:infonce-gradient}
\end{equation}

which is exactly the softmax weight of $n$ in the contrastive denominator. The gradient thus concentrates on highly similar negatives according to the model's current state, and this, without an explicit mining step. Early in training, the hardest negatives for a pre-trained encoder are topically the ones provided in the triplet, so InfoNCE allocates gradient to decorrelating style from topic, which is exactly the direction AA needs. A full derivation of Equation~\ref{eq:infonce-gradient} is given in Appendix~\ref{app:infonce-derivation}.
 
\paragraph{Multi-positive evaluation under single-positive training.}
Our training mines one positive per anchor, but at evaluation time some anchors have multiple positives, because author-sets recur across documents. Hence, in-batch ``false negatives'' (other valid positives that happen to appear in the batch and are labeled negative)  can be a concern when it comes to gradient stability. We treat these cases as benign, as in practice, occasional false negatives act like mild hard-negative noise rather than a failure mode. \citet{chen_simple_2020} report empirically that in-batch false negatives degrade performance only weakly in their self-supervised setting, and subsequent work has corroborated this. Complementarily, \cite{Khosla_supervised_2021} explicitly handles the multi-positive case and finds that aggregating multiple positives improves, rather than hurts, representation quality in supervised contrastive learning.

\subsection{Pooling and interaction}
\label{sec:method-pooling}
 
We compare three families of similarity functions, each determining how the representations produced by the encoder are combined into a document score.
 
\paragraph{Mean pooling.}
The document is represented by the mean of its token embeddings, and $s(a, p)$ is cosine similarity between the two mean vectors. This is the standard AA baseline~\citep{rivera-soto_learning_2021,huertas-tato_isolating_2024, wegmann_same_2022} and the default for general-purpose sentence encoders. We additionally trained a layerwise mean-pooling variant (with and without centering) as an alternative baseline~\citep{kantharuban_idiolex_2026}; validation-set performance only slightly outperformed mean pooling and we discarded it before test-set evaluation.
 
\paragraph{Late interaction.}
The document is represented by its full sequence of token embeddings, and $s(a, p) = \sum_{i} \max_j \cos(\mathbf{h}^a_i, \mathbf{h}^p_j)$ is the $\mathrm{MaxSim}$ score of \citet{khattab_colbert_2020}, with punctuation and padding positions masked out. This representation is used at both train and test time; the scoring function is identical to the training objective.
 
\paragraph{Patch-level late interaction (PLI).}
The token sequence is first grouped into contiguous patches, each group is mean-pooled into a single patch vector, and $\mathrm{MaxSim}$ is computed between the two resulting patch sequences. PLI reduces the per-document vector count by the average patch length and makes the granularity of interaction an explicit design choice; as a result, both the memory footprint and the $\mathrm{MaxSim}$ scoring cost drop roughly in proportion to the patch size. We study two fixed-policy families.
 
\paragraph{N-gram PLI.}
Patches are fixed-length, non-overlapping windows of $n$ tokens. We evaluate $n \in \{2, 3, 4, 5\}$. This policy is parameter-free but ignores whitespace and word boundaries, so a single linguistic word can be split across patches.
 
\paragraph{Whole-word PLI.}
Patch boundaries align with word-starts. Each patch corresponds to one linguistic word, so the representation is invariant to how the word is broken into sub-word units.
 
\paragraph{What about learned patches?}
A natural further step is to make the patch boundaries themselves trainable: a small head emits a per-token cut probability, and the boundaries are sampled with a Gumbel-Softmax straight-through estimator~\citep{jang_categorical_2017, maddison_concrete_2017} so that the patching policy can be trained jointly with the encoder under InfoNCE. We implemented this and ran it under two single-term regularizers. Both failed in characterizable ways and the resulting model never reached the fixed-$n$ baselines (Appendix~\ref{app:learned-pli}).

\section{Experiments}
\label{sec:experiments}

We evaluate each trained model with two complementary views. The first is retrieval on HALvest-Contrastive, reporting Recall@20, Recall@100, nDCG@20, and nDCG@100, with the scorer matching the training objective: cosine for mean pooling, token-level $\mathrm{MaxSim}$ for late interaction, and patch-level $\mathrm{MaxSim}$ for PLI. The second is triplet accuracy on PAN19 zero-shot.

\subsection{Topic decoupling}
\label{sec:exp-topic-decoupling}
 
The base configuration of HALvest-Contrastive is designed to prevent models from learning topical shortcuts. Before discussing any trained model, we verify empirically that the design works as intended. Two pieces of evidence support the claim.
 
\begin{figure}[h]
\centering
\includegraphics[width=\columnwidth]{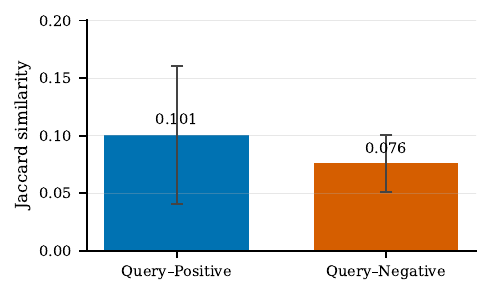}
\caption{Mean Jaccard vocabulary overlap between the components of HALvest-Contrastive triplets. Under the base configuration, query-positive overlap is already low in absolute terms, confirming that topic is decoupled between same-author passages. Query-negative overlap is lower still, but not too dissimilar.}
\label{fig:halvest_jaccard}
\end{figure}

We measure the Jaccard overlap between the word sets of triplet components on a sample of 10{,}000 triplets. Figure~\ref{fig:halvest_jaccard} reports the results.

A purely lexical retriever gives us a direct test. If the base configuration has successfully decoupled topic, BM25~\citep{robertson_okapi_1994} should succeed on the unrestricted data (where the positive can share words with the query) and fail on the base data. Table~\ref{tab:topic-decoupling} reports BM25 accuracy on both configurations across span lengths, alongside the ICT baseline and a RoBERTa~\citep{liu_roberta_2019} contrastive model trained on each configuration. Recall that we haven't scaled any configuration yet and only intend to justify our design choice for HALvest-Contrastive. On four-sentence spans, BM25 achieves 79.26\% accuracy on unrestricted but only 68.36\% on base, a 10.9-point drop. The gap narrows at longer spans, but persists throughout. The ICT model consistently outperforms BM25, especially on the base set. This indicates that its retrieval objective can learn more abstract representations. The RoBERTa contrastive model trained with triplets where the anchor and positive originate from different documents achieves top performance across all conditions.
 
\begin{table*}[t]
\centering
\small
\begin{tabular}{l rr rr rr rr}
\toprule
 & \multicolumn{2}{c}{\textbf{$k=2$}} & \multicolumn{2}{c}{\textbf{$k=4$}} & \multicolumn{2}{c}{\textbf{$k=6$}} & \multicolumn{2}{c}{\textbf{$k=8$}} \\
\cmidrule(lr){2-3}\cmidrule(lr){4-5}\cmidrule(lr){6-7}\cmidrule(lr){8-9}
\textbf{Model} & Unrestr. & Base & Unrestr. & Base & Unrestr. & Base & Unrestr. & Base \\
\midrule
BM25                      & 68.15 & 60.24 & 79.26 & 68.36 & 85.03 & 73.72 & 89.09 & 78.21 \\
ICT                       & 86.61 & 79.72 & 91.60 & 85.11 & 93.77 & 88.51 & 94.41 & 88.07 \\
RoBERTa-UnrestrData       &  ---  & 80.73 &  ---  & 87.03 &  ---  & 88.44 &  ---  & 89.91 \\
RoBERTa-BaseData          &  ---  & \textbf{82.33} &  ---  & \textbf{88.07} &  ---  & \textbf{90.15} &  ---  & \textbf{90.83} \\
\bottomrule
\end{tabular}
\caption{Accuracy (\%) on the unrestricted and base test configurations of HALvest-Contrastive across span lengths $k$. All models compared at 3.1k steps for fairness.}
\label{tab:topic-decoupling}
\end{table*}
 
We further scaled the base data afterward and made it HALvest-Contrastive.

\subsection{Effect of span length}
\label{sec:exp-span-length}
 
Accuracy rises monotonically with the span length $k$ for every model in Table~\ref{tab:topic-decoupling}but with different slopes. BM25 improves by approximately 18 percentage points on the base configuration from $k=2$ to $k=8$, or a relative gain of roughly 30\%. A lexical model is almost entirely coupled to the availability of more keywords, so doubling and then quadrupling the span length produces nearly-linear improvement. The neural contrastive models improve much more modestly: RoBERTa-BaseData gains 8.5 points (10\% relative) from $k=2$ to $k=8$, and the improvement is concentrated at the short end. The curve flattens by $k=6$.
 
Neural models seem to capture substantial authorial signal at short spans. Syntactic templates, preferred transition words, clause structures or function-word patterns are already present in two sentences. BM25, by contrast, extracts a signal that scales with how many content words happen to co-occur, a quantity that continues to grow with the span. Neural models extract a signal which marginal value saturates before the lexical signal does. All subsequent experiments use the base configuration at $k = 4$ as the primary reporting split, with full per-$k$ tables in Appendix~\ref{app:per-split}.

\subsection{Granularity across span lengths}
\label{sec:exp-granularity}
 
\begin{figure}[t]
\centering
\includegraphics[width=\columnwidth]{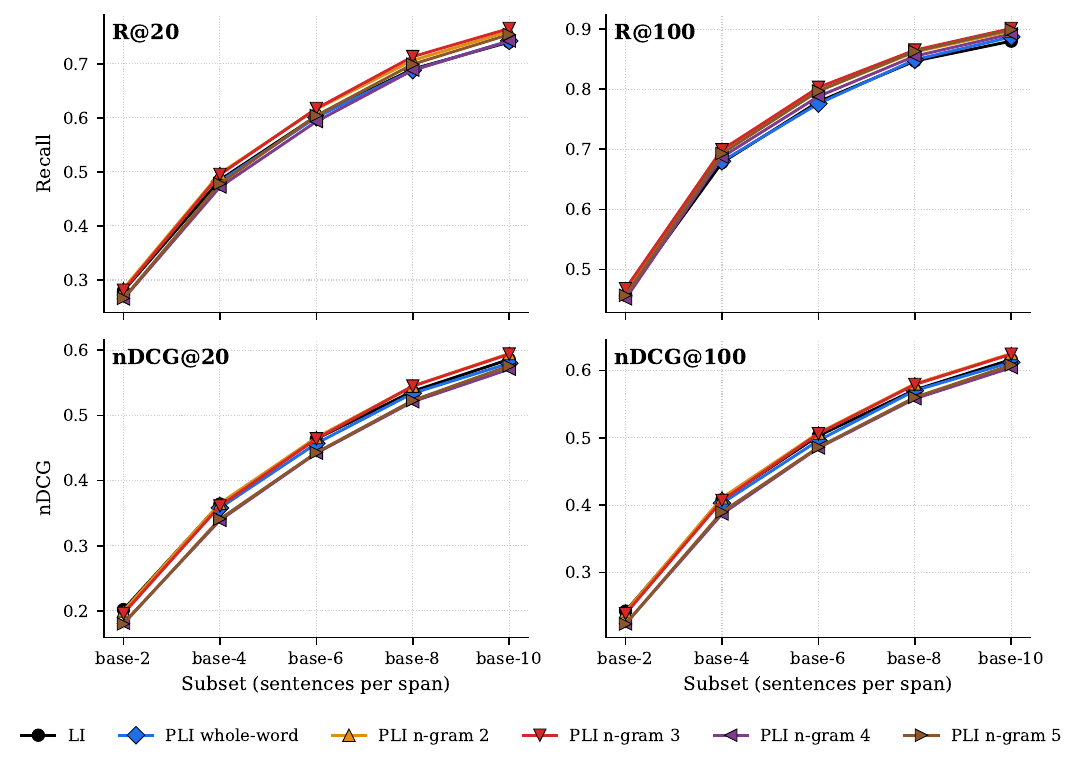}
\caption{Retrieval metrics on HALvest-Contrastive across all five subsets (base-$k$ = $k$ contiguous sentences per anchor, positive, and negative span). Each panel reports one metric, each marker family one model. Bigram patches slightly over-perform at short spans, trigram patches take over from $k$=6 onwards. The full numerical tables underlying the figure are in Appendix~\ref{app:per-split}.}
\label{fig:retrieval_metrics_by_subset}
\end{figure}

We trained ModernBERT~\citep{warner_smarter_2025} with InfoNCE for one epoch under each of the granularity introduced in \secref{sec:method-pooling}: mean pooling, late interaction, whole-word PLI, and fixed-$n$ PLI for $n \in \{2, 3, 4, 5\}$. We ran each model on every subset of HALvest-Contrastive ($k \in \{2, 4, 6, 8, 10\}$). Figure~\ref{fig:retrieval_metrics_by_subset} reports the four ranking metrics across the five subsets for the multi-vector configurations. The tables with every values are in Appendix~\ref{app:per-split}. We exclude mean pooling from the figure to keep the y-axis range readable, the comparison with mean pooling lies in a different range and is captured by the headline below.
 
Under a matched training objective, moving from single-vector pooling to late interaction yields a greater qualitative gain in AA than it yields in passage retrieval. We also observe a similar gain at inference: keeping the encoder fixed and swapping mean pooling for multi-vector scoring improves performance by about 20\% (Appendix~\ref{app:per-split}).  A similar 20\% lift appears when applying the same swap to E5, with an XLM-RoBERTa backbone~\citep{conneau_unsupervised_2020} (Appendix~\ref{app:e5-full}, Table~\ref{tab:e5-full}), suggesting the advantage is not specific to the ModernBERT backbone. Within the multi-vector family, no fixed policy is uniformly best across subsets. Token-level LI is competitive but not dominant, whole-word PLI sits in the middle of the cluster and the n-gram variants reorder as $k$ grows.
 
\subsection{Alignment and uniformity diagnostics}

\begin{figure}[h]
\centering
\includegraphics[width=\columnwidth]{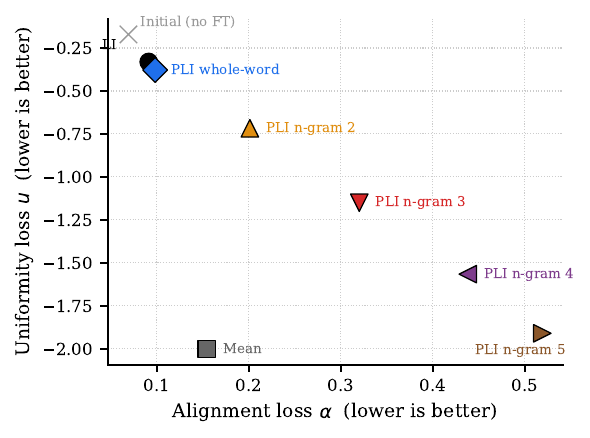}
\caption{Alignment against uniformity~\citep{wang_understanding_2020} across pooling strategies on the base-4's validation set.}
\label{fig:alignment_uniformity}
\end{figure}

We also report the alignment and uniformity losses~\citep{wang_understanding_2020} as a representation-quality diagnostic alongside retrieval and triplet metrics. Alignment measures how close positive pairs are in embedding space, while uniformity measures how spread out random pairs are.
Figure~\ref{fig:alignment_uniformity} plots the pair for every model trained in this paper.

The results align with the retrieval literature. Mean pooling has very low uniformity ($u = -2.0$) but high alignment loss ($\alpha = 0.154$): random pairs are well spread out, but same-author pairs are not particularly close. We hypothesize that mean-pooled vectors are topically well-spread, but stylistically undiscriminating, which matches explains the performance gap. LI and whole-word PLI sit in the upper-left region. The fixed-$n$ PLI variants trade alignment for uniformity along a clean monotonic axis as $n$ grows, with $n=5$: the representation has become dispersed at the cost of pulling positives apart.

\subsection{AA is not semantic retrieval}
\label{sec:exp-aa-ne-ir}
 
To test whether the gains come from semantic retrieval alone, we compute the cosine similarities between anchors and positives and between anchors and negatives using both E5~\citep{wang_text_2024} and our fine-tuned mean-pooled model on a sample of 4{,}000 triplets~\citep{kantharuban_idiolex_2026}. If the two signals were equivalent, they would separate same-author and different-author pairs in a similar way.
 
\begin{figure}[h]
\centering
\includegraphics[width=\columnwidth]{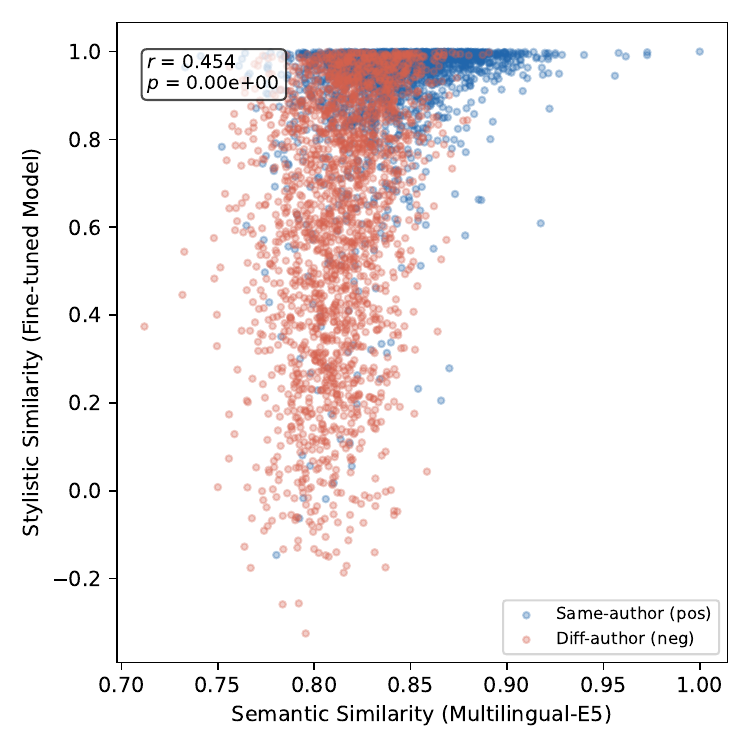}
\caption{Cosine similarity between HALvest-Contrastive anchor-candidate pairs, measured with a pretrained multilingual-E5 encoder ($x$-axis) and with our fine-tuned stylistic model ($y$-axis) using mean pooling. We report E5 performance's on HALvest-Contrastive and PAN19 in Appendix~\ref{app:e5-full}.}
\label{fig:semantic_decorrelation}
\end{figure}

Figure~\ref{fig:semantic_decorrelation} shows that neither is the case. On the stylistic axis, our fine-tuned model separates same-author pairs from different-author pairs by a margin of 0.30 in mean cosine similarity. On the semantic axis, E5 separates the same populations by only 0.029. Semantic similarity under E5 is therefore almost invariant to the authorship relation between two passages, whereas the stylistic representation responds to it strongly. The pooled Pearson correlation across all 4{,}000 pairs is $r = 0.454$, which may suggest partial agreement between the two signals. This coefficient, however, is dominated by the between-group shift we just described: the within-group structure in the scatter shows stylistic similarity varying across the full range while semantic similarity stays confined in a narrow vertical band.
 
\subsection{Retrieval failure modes}
\label{sec:exp-failure-modes}
 
To further emphasize the distinction between AA and simple retrieval, we inspect the failure modes of our strongest  PLI baseline (PLI n-gram 2 on $k=4$) by ranking 500 sampled queries against a 1000-document validation pool (5 seeds $\times$ 100 queries). The model places the true positive at rank 1 in 20.8\% of cases and within the top 5 in 39.8\%, with a long tail (16.6\% of true positives at ranks 101--500 and 8.0\% beyond rank 500).
 
We call ``distractors'' highly ranked negative documents, and categorize them. Some distractors share the query's domain at 61\% across rank buckets, against 20\% for random pairs. The model still has learned domain as a strong similarity cue, consistent with HALvest-Contrastive's field-matched negative sampling. The top-K distractors share at least one author with the query in 11.4\% of cases at ranks 2--5 and 11.8\% at ranks 6--20, but 0.0\% across 500 random pairs. The author-overlap drops by a factor of 3.5 between the top-20 distractors and the subsequent ones ranked lower. The presence of overlapping authors correlates strongly with retrieval success. The Jaccard distance with the query (unigram) is essentially flat across rank buckets (0.086, 0.087, 0.085, with random at 0.073). The Jaccard distance with the true positives however, is more informative: 0.129 at rank 1, 0.131 at ranks 2--5, 0.126 at ranks 6--20, dropping to 0.076 at lower ranks. High-ranked distractors do not look textually similar to the query, but they do look textually similar to the true positive. This is structurally different from a topic-retrieval failure mode, where lexical query overlap would dominate.

\subsection{Cross-domain and patch-size scaling}
\label{sec:exp-pan19}

\begin{table}[ht]
\centering
\small
\begin{tabular}{lr}
\toprule
\textbf{Model} & \textbf{Triplet accuracy} \\
\midrule
Mean pooling        & 0.520 \\
LI (full token)     & 0.719 \\
PLI whole-word      & 0.715 \\
PLI n-gram 5        & 0.704 \\
PLI n-gram 3        & 0.724 \\
PLI n-gram 2        & \textbf{0.740} \\
PLI n-gram 4        & \textbf{0.740} \\
\midrule
E5 zero-shot        & 0.628 \\
\bottomrule
\end{tabular}
\caption{Zero-shot triplet accuracy on PAN19. All trained models are evaluated without any further fine-tuning.}
\label{tab:pan19}
\end{table}

Table~\ref{tab:pan19} reports triplet accuracy on PAN19 for every trained model, evaluated zero-shot. The pattern within fixed-policy variants is consistent with what we saw on HALvest-Contrastive's longer subsets: token-level LI is competitive but not best, whole-word sits in the middle.
 
\paragraph{An empirical patch-size fit.}
The shift in optimal $n$ across subsets is well approximated by a one-parameter square-root fit. Let $S$ denote the token count of an anchor, which varies by subset because longer-$k$ subsets fill more of the ModernBERT context (512 tokens). Across six subsets with span length ranging from 69.5 tokens to 512.0 tokens, the best fixed $n$ is fitted by
 
\begin{equation}
n_\star \;\approx\; 0.18 \cdot \sqrt{S}.
\label{eq:sqrt-law}
\end{equation}
 
\begin{table}[h]
\centering
\small
\begin{tabular}{lrrr}
\toprule
\textbf{Subset} & \textbf{$S$} & $0.18 \sqrt{S}$ & $n_\star$ \\
\midrule
HALvest-C base-2   &  69.5 & 1.50 & 1 \\
HALvest-C base-4   & 133.2 & 2.08 & 2 \\
HALvest-C base-6   & 201.4 & 2.55 & 3 \\
HALvest-C base-8   & 266.2 & 2.94 & 3 \\
HALvest-C base-10  & 327.6 & 3.26 & 3 \\
PAN19 (zero-shot)  & 512.0 & 4.07 & 4 \\
\bottomrule
\end{tabular}
\caption{Empirical fit of $n_\star$. At PAN19's sequence length the fit matches one of the two tied values. The limited resolution of our patch sizes means ties of this kind are expected.}
\label{tab:sqrt-law}
\end{table}
 
The relationship is empirical, not derived. The relationship also explains the apparent disagreement between rankings in Figure~\ref{fig:retrieval_metrics_by_subset} for a given subset: each subset samples a different point on the same curve.. We treat Equation \eqref{eq:sqrt-law} as a descriptive fit over the range we tested. It tells a practitioner who has chosen a context length what fixed patch size to start at, but does not say why $\sqrt{\cdot}$ is the right exponent. Under that empirical fit, the best-performing patch size grows sublinearly with sequence length, implying that the number of stored patch vectors also grows sublinearly (approximately proportional to $\sqrt{S}$ over the range we tested). Therefore patch-level interaction gives a much cheaper alternative to full token-level LI while preserving most of the gain on AA.

\subsection{Scaling the negative-pool size}
\label{sec:exp-scaling}
 
To verify that HALvest-Contrastive exhibits the same qualitative scaling as retrieval benchmarks, we retrained the late-interaction model at per-device batch 64, yielding 512 in-batch negatives per anchor after all-gather (versus 256 at batch 32). Retrieval metrics improved by a small but consistent margin on base-4 and the neighboring splits. R@20 increases by a fraction of a percentage point, R@100 by slightly more, and the direction is the same across nDCG metrics. This is consistent with the standard retrieval findings (\secref{sec:related}).

\section{Conclusion}
\label{sec:conclusion}

Scholarly prose is a strong testbed for stylometry because its low-entropy register compresses topical variation and makes the stylistic cues easier to isolate. HALvest-Contrastive exploits this structure by pairing same-author passages across different papers within a disciplinary field. Our large-scale benchmark substantially weakens lexical shortcuts.

The transfer from retrieval to authorship attribution is not uniform across design choices. Moving from single-vector pooling to multi-vector interaction produces large gains, but full token-level matching is not uniformly optimal. Grouping neighboring tokens into patches preserves most of the retrieval improvement while reducing the number of stored vectors and pairwise $\mathrm{MaxSim}$ comparisons. Across both HALvest-Contrastive and PAN19, the best-performing fixed patch size increases gradually with sequence length and is well approximated, over the range we tested, by $n_\star \approx 0.18 \cdot \sqrt{S}$.

More broadly, our results suggest that authorship attribution behaves differently from semantic retrieval despite their shared contrastive structure. Semantic matching benefits from fine-grained lexical alignment, whereas stylometric similarity appears to emerge at an intermediate interaction scale.

\section*{Limitations}
\label{sec:limitations}
 
\paragraph{Author-set rather than individual attribution.}
The triplet label is the set of co-authors rather than a single writer. We argue in \secref{sec:dataset} that this is the correct target for scholarly repositories and that the forensic-linguistics literature supports it as a principled design choice rather than a concession~\citep{dauber_git_2019}. Author-set labels represent a reliable level of supervision already available in scholarly metadata. Individual attribution within multi-author documents is a different task, and we leave it to future work.
 
\paragraph{GROBID.}
HALvest's text is produced by running GROBID over PDFs. Relying on GROBID as a proxy before performing PDF parsing may seem indirect, as more accurate ways exist to extract text from PDFs---vision-language models and Optical Character Recognition (OCR), for instance. However, having a structured view of our articles enables finer control over text output. GROBID’s layout-based extraction preserves document structure, essential for our contrastive sampling strategy where we need to identify and extract contiguous sentence spans across different sections of papers. GROBID, despite being a conservative extractor, can introduce systematic biases. First, Arabic scripts are inverted in the output: word order is reversed because the extractor's layout model assumes left-to-right reading, which makes the Arabic portion of HALvest effectively unusable. Second, CJK languages lack the whitespace cues that GROBID's tokeniser relies on for length-based filtering, so coverage in these languages is sparser and more variable. Third, mathematical notation is often stripped or rendered inconsistently, which can bias mathematical-sciences domains away from passages where notation is central. As a rough estimation, preprocessing of HAL's French split at the sentence level~\citep{antoun_camembert_2024} retained about 52\% of the total tokens (4.7 billion tokens) after tab removal and formula handling, and this retention rate varies across languages.
 
\paragraph{Learned patching is unfinished.}
We attempted an end-to-end trainable patching policy but found that simple regularizers either collapse the policy to token-level scoring (when the regularizer is unbounded below) or to single-patch sequences (when constraining only the population mean of cut probabilities below the inference threshold). See Appendix~\ref{app:learned-pli} for the math and the empirical failure-mode characterization. A successful learned variant would require multi-term regularization in the BLT~\citep{pagnoni_byte_2025} style with hyperparameter tuning beyond our compute budget for this submission. The cross-attention compressor variant of \citet{pagnoni_byte_2025} was explored alongside it and shares the same regularizer-design weaknesses.

\paragraph{Empirical scaling fit.}
The $0.18 \cdot \sqrt{S}$ relationship in \secref{sec:exp-pan19} is descriptive.  Six points across two registers fit a one-parameter curve cleanly, but we do not provide a derivation, and we do not yet know whether the same coefficient transfers to other encoders or longer context lengths. It should be read as a practical summary for choosing a starting patch size, not as a universal law.

\section*{Ethics}
\label{sec:ethics}
 
\paragraph{Licensing.}
HAL is an open-access repository. The documents in HALvest were selected from papers deposited under permissive licenses compatible with derivative redistribution.
 
\paragraph{Author identity.}
The author labels in HALvest are exactly those that appear in the public metadata of the source papers. No personally identifiable information beyond what is already published is introduced by the derivative corpus. Author names are tied to signed, publicly-archived academic works; the corpus does not include private correspondence or unpublished drafts.
 
\paragraph{Dual-use considerations.}
Stylometric authorship attribution is a dual-use technology. It enables beneficial applications (plagiarism detection, LLM-generated-text detection, verification of anonymous peer review integrity) but can in principle be used to deanonymize pseudonymous writers, including whistleblowers and politically vulnerable authors. HALvest itself contains only signed academic work, so the corpus-specific deanonymisation risk is low; however, models trained on HALvest may be portable, and we encourage users deploying HALvest-trained models in downstream tasks to consider the risk surface of their own application domain.

\section*{Acknowledgments}

The authors are grateful to the CCSD staff—the service unit in charge of HAL—in particular Achraf Azhar. We also thank Patrice Lopez, for providing resources and support to better handle GROBID, as well as Arij Riabi, Brahim Talb and Menel Mahamdi for the productive discussions. Finally, we extend our thank to Yannis Karmim and Lydia Nishimwe for their proofreading. This work was partially realized on computing HPC and storage resources provided by IDRIS thanks to the grant GCDA1016807 on the DALIA supercomputer.

\bibliography{custom}

\appendix

\section{InfoNCE gradient derivation}
\label{app:infonce-derivation}
 
We derive Equation~\ref{eq:infonce-gradient} from Equation~\ref{eq:infonce}. Writing $Z = \exp(s_p / \tau) + \sum_{n' \in \mathcal{N}} \exp(s_{n'} / \tau)$ for the partition function and $s_p = s(a, p)$, $s_n = s(a, n)$ for the anchor-positive and anchor-negative similarities, the InfoNCE loss for a single anchor is
\[
\mathcal{L} = -\log \frac{\exp(s_p / \tau)}{Z} = -\frac{s_p}{\tau} + \log Z.
\]
Differentiating with respect to a specific negative similarity $s_n$ (treating $s_p$ and the other negatives $\{s_{n'} : n' \neq n\}$ as fixed inputs),
 
\begin{align*}
\frac{\partial \mathcal{L}}{\partial s_n}
&= \frac{1}{Z} \cdot \frac{1}{\tau} \cdot \exp(s_n / \tau) \\
&= \frac{1}{\tau} \cdot \frac{\exp(s_n / \tau)}{\exp(s_p / \tau) + \sum_{n' \in \mathcal{N}} \exp(s_{n'} / \tau)}.
\end{align*}
 
The right-hand side is exactly the softmax weight of $n$ in the contrastive denominator, scaled by $1 / \tau$. Negatives with high similarity to the anchor receive exponentially more gradient than negatives with low similarity. In the limit of one dominant negative, the gradient collapses onto that negative alone, reproducing the behaviour of hard-negative mining without an explicit mining procedure.
 
The comparison with the triplet loss follows from inspecting its form directly. With a fixed margin $m$, the triplet loss is $\mathcal{L}_{\mathrm{trip}}(s_p, s_n) = [m - s_p + s_n]_+$, whose gradient with respect to $s_n$ is $+1$ when $s_p - s_n < m$ and $0$ otherwise. The gradient does not depend on how close $s_n$ is to $s_p$ within the active region, and it vanishes entirely once the margin is satisfied. In that sense, triplet loss does not have a hard-negative-focus property in the InfoNCE sense: the loss treats all active negatives identically and ignores all inactive ones.

\section{HALvest language and domain statistics}
\label{app:hal-stats}
 
\begin{table}[h]
\centering
\small
\begin{tabular}{lrr}
\toprule
\textbf{Language} & \textbf{\# Docs} & \textbf{\# Tokens} \\
\midrule
English       & 464{,}679 & 8{,}158{,}933{,}235 \\
French        & 199{,}216 & 9{,}018{,}529{,}985 \\
Spanish       &   2{,}975 &    69{,}221{,}667 \\
Italian       &   1{,}172 &    48{,}747{,}986 \\
Portuguese    &      934 &    32{,}918{,}832 \\
German        &      652 &    12{,}225{,}960 \\
Russian       &      245 &     5{,}763{,}532 \\
Chinese       &      160 &     2{,}861{,}585 \\
\textit{\ldots}      &     \textit{\ldots}      &     \textit{\ldots}    \\
\midrule
\textbf{Total (56 languages)} & \textbf{778k+} & \textbf{17.4B} \\
\bottomrule
\end{tabular}
\caption{Language distribution of HALvest. Full per-language counts (56 languages, including Basque, Catalan, Persian, and other low-resource entries) are included with the dataset release.}
\label{tab:hal-languages}
\end{table}
 
\begin{table}[h]
\centering
\tiny
\begin{tabular}{llrr}
        \toprule
        \textbf{Domain} & \textbf{Code} & \textbf{\# Documents} & \textbf{\# Tokens} \\
        \midrule
        Humanities and Social Sciences & shs & 156,566 & 5,614,423,171 \\
        Computer Science & info & 148,316 & 2,573,673,455 \\
        Life Sciences & sdv & 115,744 & 3,145,323,780 \\
        Engineering Sciences & spi & 102,751 & 2,254,653,825 \\
        Physics & phys & 65,991 & 1,503,190,749 \\
        Mathematics & math & 62,921 & 1,638,500,361 \\
        Chemical Science & chim & 40,012 & 899,507,319 \\
        Environmental Science & sde & 31,575 & 579,076,669 \\
        Sciences of the Universe & sdu & 23,557 & 682,356,264 \\
        Cognitive Science & scco & 11,772 & 227,487,096 \\
        Statistics & stat & 10,579 & 184,678,350 \\
        Quantitative Finance & qfin & 3,451 & 68,518,636 \\
        Nonlinear Sciences & nlin & 1,972 & 30,694,088 \\
        \bottomrule
    \end{tabular}
\caption{Disciplinary domains retained for HALvest-Contrastive hard-negative mining. Three of the 16 HALvest domains are excluded from the contrastive derivation for having too few documents to support field-matched negative sampling; they remain in HALvest itself.}
\label{tab:hal-domains}
\end{table}
 
We selected eight features to represent HAL submissions:
 
\begin{itemize}
\item \texttt{halid}: submission's unique identifier assigned by HAL.
\item \texttt{lang}: the language of the document, as filled by the depositor.
\item \texttt{title}: title of the document.
\item \texttt{domain}: list of fields of study\footnote{\url{https://hal.science/browse/domain}}.
\item \texttt{timestamp}: time of access.
\item \texttt{year}: publication year of the document if relevant. Otherwise, it is set to year 1.
\item \texttt{url}: URL to access the \textsc{pdf}.
\item \texttt{authors}: list of authors.
\end{itemize}
 
Table~\ref{tab:hal-languages} reports per-language document and token counts for HALvest. Table~\ref{tab:hal-domains} reports the 13 domains retained for HALvest-Contrastive triplet mining.
 
\begin{figure}[h]
\centering
    \includegraphics[width=\columnwidth]{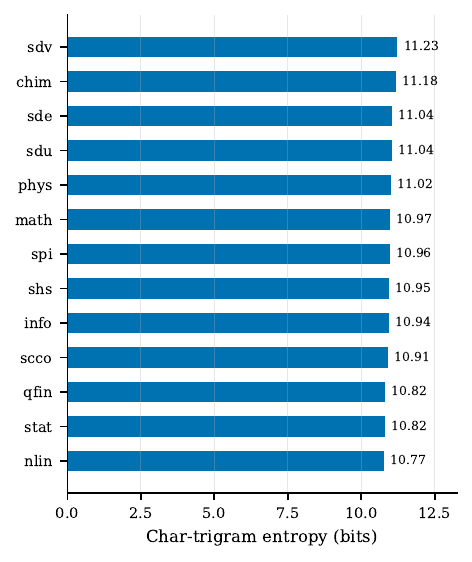}
    \caption{Token-level entropy across the 13 disciplinary domains of HALvest.}
\label{fig:halvest_entropy}
\end{figure}
 
Figure~\ref{fig:halvest_entropy} plots token-level entropy across domains, which is approximately uniform. Academic writing remains low-entropy throughout the fields.
 
\begin{figure}[h]
\centering
\includegraphics[width=\columnwidth]{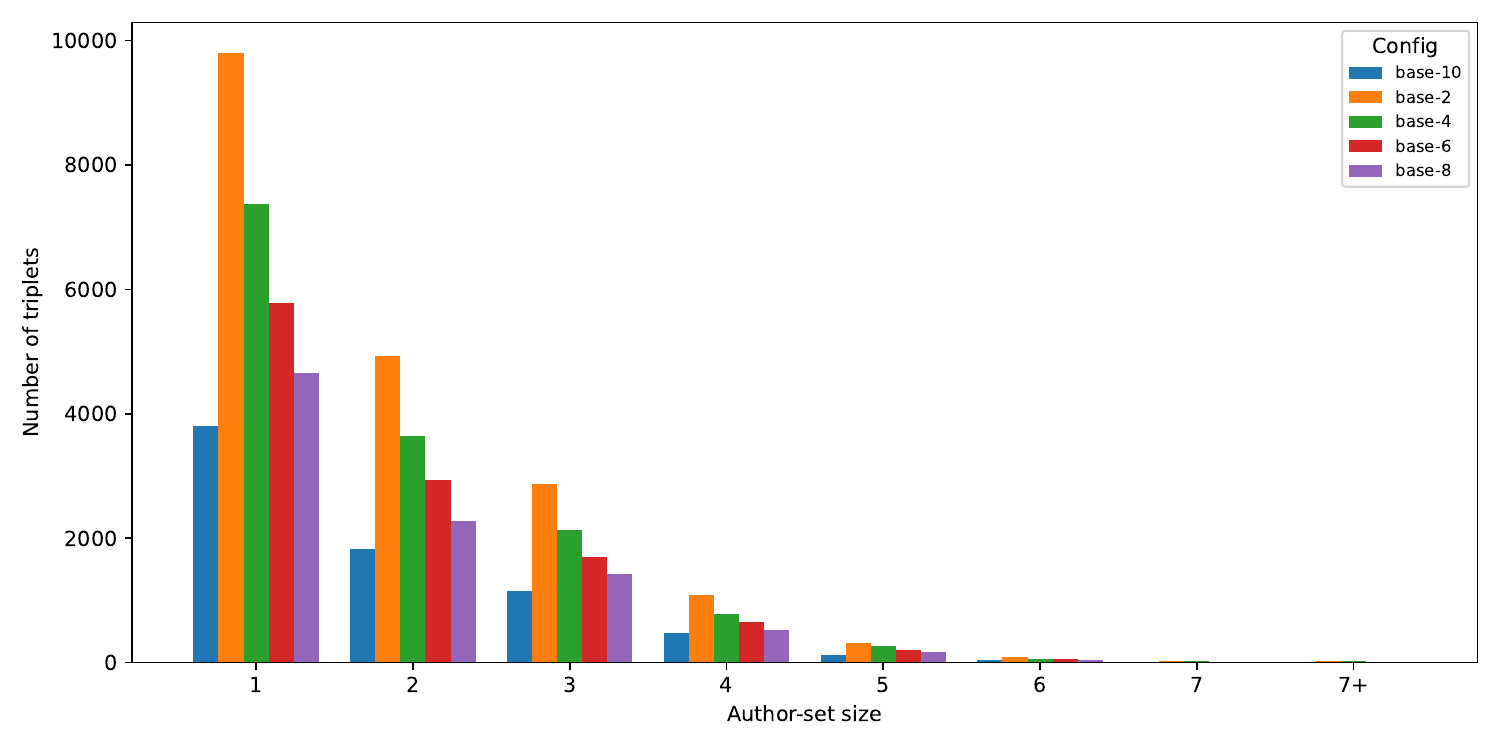}
\caption{Author-set sizes in HALvest-Contrastive. Each color represents a subset of $k$ contiguous sentences. Respectively 10, 2, 4, 6 and 8. Subsets are grouped by their number of author in ascending order. The majority of the documents, no matter the subset are single-authored.}
\label{fig:collaboration_recurrence}
\end{figure}
 
Figure~\ref{fig:collaboration_recurrence} plots the author sets' distribution. The median author-set size is 1, the mean is 1.84, and the maximum set size in the corpus is 44.
 
An author comprises three features:
 
\begin{itemize}
\item \texttt{name}: string for the author name, as filled by the depositor.
\item \texttt{affiliations}: list of unique identifiers attributed by HAL to the institutions where the author belongs.
\item \texttt{halauthorid}: unique identifier assigned by HAL to each author registered on the online repository. If an author is not registered on HAL, they are considered unidentified and are assigned a \texttt{halauthorid} of ``0''.
\end{itemize}

\section{Full hyperparameters and setup}
\label{app:hyperparams}
 
Unless stated otherwise, all trained models share a ModernBERT-base backbone~\citep{warner_smarter_2025} and are trained for a single epoch on the base configuration of HALvest-Contrastive. Experiments use four H100 GPUs with a per-device batch size of 32 and gradient accumulation of 2, so that the effective batch is 64 and each anchor is contrasted against 256 in-batch negatives after all-gather. The learning rate is $3 \times 10^{-5}$, the InfoNCE temperature $\tau$ is 0.5, weight decay is 0.1, and we train in 16-bit mixed precision. Runs with four GB200 at a per-device batch size of 64 (512 in-batch negatives) are reported in \secref{sec:exp-scaling} as a scaling reference.
 
\begin{table}[h]
\centering
\tiny
\begin{tabular}{lc}
\toprule
\textbf{Hyperparameter} & \textbf{Value} \\
\midrule
Backbone & ModernBERT-base \\
Epochs & 1 \\
Per-device batch size & 32 \\
Gradient accumulation & 2 \\
GPUs & 4$\times$H100 (primary), 4$\times$GB200 (scaling) \\
Effective in-batch negatives & 256 (H100), 512 (GB200) \\
Learning rate & $3 \times 10^{-5}$ \\
Weight decay & 0.1 \\
Precision & 16-bit mixed \\
InfoNCE temperature $\tau$ & 0.5 \\
\bottomrule
\end{tabular}
\caption{Training and model hyperparameters shared across all main-body models. Learned-PLI hyperparameters (kept for reproducibility despite the negative result) appear in Appendix~\ref{app:learned-pli}.}
\label{tab:hyperparams}
\end{table}
 
Table~\ref{tab:hyperparams} reproduces the training configuration of every main-body model. $\tau$ is a temperature parameter, set to 0.5 throughout. This value accounts for the large magnitude of $\mathrm{MaxSim}$ scores on long sequences, which require a higher temperature than typical passage-retrieval settings ($\tau \in \left[0.05, 0.1\right]$) to avoid saturating the softmax. All other settings are defaults of the PyTorch Lightning~\citep{Falcon_PyTorch_Lightning_2019} training pipeline that accompanies the release.

\section{Per-split results}
\label{app:per-split}
 
Tables~\ref{tab:split-base2} to~\ref{tab:split-base10} report retrieval metrics for every trained model on every test split (base-2, base-4, base-6, base-8, base-10), including the learned-PLI variants whose negative-result analysis is in Appendix~\ref{app:learned-pli}. We omit layerwise pooling variants, abandoned after validation-set evaluation only marginally improving over mean pooling.
 
\begin{table}[h]
\centering
\scriptsize
\begin{tabular}{lrrrr}
\toprule
\textbf{Model}            & \textbf{R@20} & \textbf{R@100} & \textbf{nDCG@20} & \textbf{nDCG@100} \\
\midrule
        Mean pooling            & 0.060         & 0.173         & 0.032           & 0.058            \\
        Mean pooling (LI)       & 0.073         & 0.190         & 0.041           & 0.067 \\
        LI                      & 0.280         & 0.457         & \textbf{0.202}           & \textbf{0.243}            \\
        PLI n-gram 2            & \textbf{0.284}         & 0.465          & 0.200            & 0.242             \\
        PLI n-gram 3            & 0.281         & \textbf{0.468}          & 0.196            & 0.239             \\
        PLI n-gram 4            & 0.265         & 0.451          & 0.181            & 0.224             \\
        PLI n-gram 5            & 0.266         & 0.457          & 0.180            & 0.224             \\
        PLI whole-word          & --            & --             & --               & --                \\
        PLI learned             & 0.258            & 0.450             & 0.171               & 0.215                \\
\bottomrule
\end{tabular}
\caption{HALvest-Contrastive \textbf{base-2}: primary retrieval metrics. Learned-PLI rows reflect the retrained checkpoint analysed in Appendix~\ref{app:learned-pli}; the policy collapses to single-patch sequences and the resulting numbers track mean pooling.}
\label{tab:split-base2}
\end{table}
 
\begin{table}[h]
\centering
\scriptsize
\begin{tabular}{lrrrr}
\toprule
\textbf{Model}            & \textbf{R@20} & \textbf{R@100} & \textbf{nDCG@20} & \textbf{nDCG@100} \\
\midrule
        Mean pooling            & 0.121         & 0.294          & 0.063            & 0.101             \\
        Mean pooling (LI)       & 0.147         & 0.328          & 0.082            & 0.121              \\
        LI                      & 0.485         & 0.678          & 0.364            & 0.408             \\
        PLI n-gram 2            & \textbf{0.497}         & \textbf{0.700}          & \textbf{0.365}            & \textbf{0.411}             \\
        PLI n-gram 3            & 0.495         & 0.699          & 0.361            & 0.407             \\
        PLI n-gram 4            & 0.472         & 0.687          & 0.339            & 0.387             \\
        PLI n-gram 5            & 0.478         & 0.693          & 0.341            & 0.390             \\
        PLI whole-word          & 0.482         & 0.680          & 0.358            & 0.403             \\
        PLI learned             & 0.452         & 0.670          & 0.314            & 0.363                \\
\bottomrule
\end{tabular}
\caption{HALvest-Contrastive \textbf{base-4}: primary retrieval metrics.}
\label{tab:split-base4}
\end{table}
 
\begin{table}[h]
\centering
\scriptsize
\begin{tabular}{lrrrr}
\toprule
\textbf{Model}            & \textbf{R@20}  & \textbf{R@100} & \textbf{nDCG@20} & \textbf{nDCG@100} \\
\midrule
        Mean pooling            & 0.163        & 0.377         & 0.087           & 0.132            \\
        Mean pooling (LI)       & 0.193         & 0.420          & 0.107            & 0.156         \\
        LI                      & 0.603        & 0.779         & 0.464           & 0.503            \\
        PLI n-gram 2            & 0.616         & 0.797          & \textbf{0.466}            & \textbf{0.507}             \\
        PLI n-gram 3            & \textbf{0.617}         & \textbf{0.803}          & 0.464            & 0.506             \\
        PLI n-gram 4            & 0.593         & 0.788          & 0.442            & 0.485             \\
        PLI n-gram 5            & 0.604         & 0.797          & 0.443            & 0.486             \\
        PLI whole-word          & 0.601        & 0.776         & 0.457           & 0.496            \\
        PLI learned             & --            & --             & --               & --                \\
\bottomrule
\end{tabular}
\caption{HALvest-Contrastive \textbf{base-6}: primary retrieval metrics.}
\label{tab:split-base6}
\end{table}
 
\begin{table}[h]
\centering
\scriptsize
\begin{tabular}{lrrrr}
\toprule
\textbf{Model}            & \textbf{R@20}  & \textbf{R@100} & \textbf{nDCG@20} & \textbf{nDCG@100} \\
\midrule
        Mean pooling            & 0.211        & 0.450         & 0.114           & 0.164            \\
        Mean pooling (LI)       & 0.247         & 0.494          & 0.143            & 0.195         \\
        LI                      & 0.690        & 0.847         & 0.537           & 0.571            \\
        PLI n-gram 2            & 0.706         & 0.862          & \textbf{0.545}            & \textbf{0.580}             \\
        PLI n-gram 3            & \textbf{0.713}         & \textbf{0.865}          & \textbf{0.545}            & 0.579             \\
        PLI n-gram 4            & 0.688         & 0.855          & 0.521            & 0.558             \\
        PLI n-gram 5            & 0.699         & 0.863          & 0.523            & 0.560             \\
        PLI whole-word          & 0.688        & 0.849         & 0.534           & 0.570            \\
        PLI learned             & 0.659            & 0.838             & 0.483               & 0.522                \\
\bottomrule
\end{tabular}
\caption{HALvest-Contrastive \textbf{base-8}: primary retrieval metrics.}
\label{tab:split-base8}
\end{table}
 
\begin{table}[h]
\centering
\scriptsize
\begin{tabular}{lrrrr}
\toprule
\textbf{Model}            & \textbf{R@20}  & \textbf{R@100} & \textbf{nDCG@20} & \textbf{nDCG@100} \\
\midrule
        Mean pooling            & 0.246        & 0.499         & 0.130           & 0.182            \\
        Mean pooling (LI)       & 0.287         & 0.547          & 0.163            & 0.217         \\
        LI                      & 0.740        & 0.880         & 0.586           & 0.616            \\
        PLI n-gram 2            & 0.760         & 0.897          & \textbf{0.595}            & \textbf{0.625}             \\
        PLI n-gram 3            & \textbf{0.765}         & \textbf{0.901}          & 0.594            & 0.624             \\
        PLI n-gram 4            & 0.742         & 0.892          & 0.571            & 0.604             \\
        PLI n-gram 5            & 0.754         & 0.899          & 0.576            & 0.608             \\
        PLI whole-word          & 0.742        & 0.887         & 0.580           & 0.612            \\
        PLI learned             & 0.714            & 0.872             & 0.534               & 0.572                \\
\bottomrule
\end{tabular}
\caption{HALvest-Contrastive \textbf{base-10}: primary retrieval metrics.}
\label{tab:split-base10}
\end{table}

\section{Alignment and uniformity values}
\label{app:align-unif}
 
Numerical values for Figure~\ref{fig:alignment_uniformity} are in Table~\ref{tab:align-unif}.
 
\begin{table}[h]
\centering
\small
\begin{tabular}{lcc}
\toprule
\textbf{Model} & $\alpha$ & $u$ \\
\midrule
Pretrained, no fine-tuning & 0.069 & $-0.171$ \\
Mean pooling               & 0.154 & $-2.000$ \\
LI (full token)            & 0.091 & $-0.331$ \\
PLI whole-word             & 0.098 & $-0.378$ \\
PLI n-gram 2               & 0.201 & $-0.717$ \\
PLI n-gram 3               & 0.320 & $-1.150$ \\
PLI n-gram 4               & 0.438 & $-1.565$ \\
PLI n-gram 5               & 0.519 & $-1.909$ \\
\bottomrule
\end{tabular}
\caption{Alignment $\alpha$ and uniformity $u$ per model. Lower $\alpha$ indicates positives are closer in embedding space. Lower $u$ indicates that random pairs are more uniformly distributed. The learned-PLI row reflects the retrained checkpoint analysed in Appendix~\ref{app:learned-pli}.}
\label{tab:align-unif}
\end{table}

\section{Learned PLI: failure-mode analysis}
\label{app:learned-pli}
 
This appendix documents our attempt at learning differentiable patch boundaries. The architecture is rather straightforward. The difficulty lies in the regularizer. We tried two natural single-term forms. The first not lower bounded and collapsed the policy onto token-level scoring. The second was bounded but failed to interact with the inference threshold and collapsed the policy onto a single patch per sequence. A successful regularizer would ressemble that of BLT~\citep{pagnoni_byte_2025} but require hyperparameter sweep beyond our compute budget for this article.

\subsection{Architecture and notation}
\label{app:learned-pli-arch}
 
Let $\mathbf{h}_t \in \mathbb{R}^d$ be the encoder's hidden state at token position $t$, and let $S$ denote the number of valid (non-padding) query positions in a sequence. A two-layer feed-forward network (a patch-boundary predictor) on top of the ModernBERT hidden states produces a boundary logit per position,

\begin{equation}
\ell_t = W_2 \, \phi(W_1 \mathbf{h}_t + \mathbf{b}_1) + b_2,
\label{eq:boundary-logit}
\end{equation}

with $W_1 \in \mathbb{R}^{(d/4) \times d}$, $W_2 \in \mathbb{R}^{1 \times (d/4)}$, and $\phi$ a GELU. The cut probability is $q_t = \sigma(\ell_t)$. At training time, cut decisions are sampled with a Gumbel-Softmax alongside a straight-through gradient estimation. We anneal the temperature $\tau_{\mathrm{GS}}$ from $1.0$ to $0.1$ over the first $20{,}000$ optimizer steps. The first valid position is forced to be a cut, this guarantees at least one patch regardless of the sampler. At inference time, the cut decision is hard-thresholded at $0.5$.

The total training loss is $\mathcal{L} = \mathcal{L}_{\mathrm{InfoNCE}} + \lambda \cdot \mathcal{L}_{\mathrm{patch}}$ with $\lambda = 0.1$, where $\mathcal{L}_{\mathrm{patch}}$ is the patch regularizer whose role is to prevent degenerate solutions.
 
\subsection{Unbounded log-sum penalty}
\label{app:learned-pli-r1}
 
The first form was

\begin{equation}
\mathcal{L}^{(1)}_{\mathrm{patch}} \;=\; -\log\!\Big(\sum_{t=1}^{S} q_t\Big).
\label{eq:reg1}
\end{equation}

The intuition was: as $\sum_t q_t \to 0$, the regularizer diverges to $+\infty$ and rules out the single-patch degeneracy. Beyond that, the regularizer was supposed to be roughly flat at high cut counts, letting InfoNCE push the cut count freely.
 
However, for $\sum_t q_t > 1$, the function $-\log(\sum_t q_t)$ is monotonically decreasing in $\sum_t q_t$ and is not lower bounded. The regularizer, therefore, supplies a uniform downward push on every cut probability, regardless of position.
 
What \eqref{eq:reg1} needed was a regularizer that diverges to $+\infty$ at \emph{both} extremes of $\sum_t q_t$. Restoring the missing one requires committing to a target rate, which is what the second attempt did.

\subsection{Target-rate squared error}
\label{app:learned-pli-r2}
 
The second form was a target-rate-matching squared-error penalty,

\begin{equation}
\mathcal{L}^{(2)}_{\mathrm{patch}} \;=\; \Big(\frac{1}{S}\sum_{t=1}^{S} q_t \;-\; p_{\text{target}}\Big)^2,
\label{eq:reg2}
\end{equation}

with $p_{\text{target}} = 1/3$. This target rate corresponds to a mean patch length of $1/p_{\text{target}} = 3$ tokens, which lands in the middle of the empirical $n$-gram optima we identified (best $n$ is 2 at base-2 and base-4, 3 at base-6 onwards). The hyperparameter is therefore not tuned but set from independent empirical evidence. Equation \eqref{eq:reg2} has a minimum at $\bar{q} \coloneqq \tfrac{1}{S}\sum_t q_t = p_{\text{target}}$, is lower-bounded at zero, is symmetric and is differentiable everywhere.
 
\paragraph{The retrained model satisfies the regularizer exactly} and produces a single patch per sequence.
After retraining under $\mathcal{L}^{(2)}_{\mathrm{patch}}$, the diagnostic check on the resulting checkpoint reveals $\bar{q} \approx 1/3$ for every sequence: $\mathcal{L}^{(2)}_{\mathrm{patch}}$ is doing exactly what was asked of it. However, cut-off the threshold is $0.5$. If every position has $q_t \approx 1/3$, then no position satisfies $q_t > 0.5$, and the inference rule produces zero cuts.
 
The MSE regularizer only constrains the population mean of cut probabilities. It does not place any constraint on the distribution shape. Besides, InfoNCE provides no pressure to break the uniform policy. The regularizer does not align its constraint with the inference threshold. With $p_{\text{target}} = 1/3$ and threshold $0.5$, the model can satisfy the regularizer exactly while producing zero cuts at inference. Setting $p_{\text{target}} = 0.5$ does not fix this, as $q_t = 0.5$ everywhere produces random cuts under Gumbel sampling.
 
The cleanest fix follows the byte-latent-transformer pattern~\citep{pagnoni_byte_2025}, two regularizers acting in concert. Both regularizers introduce hyperparameters that need joint tuning.

\section{E5 zero-shot full metrics}
\label{app:e5-full}
 
\begin{table}[h]
\centering
\small
\begin{tabular}{lcc}
\toprule
\textbf{Metric} & \textbf{Dense} & \textbf{LI} \\
\midrule
\multicolumn{3}{l}{\textit{HALvest-Contrastive base-4}} \\
R@20            & 0.167 & 0.203 \\
R@100           & 0.269 & 0.307 \\
nDCG@20         & 0.124 & 0.152 \\
nDCG@100        & 0.146 & 0.175 \\
Triplet accuracy & \multicolumn{2}{c}{0.789} \\
\midrule
\multicolumn{3}{l}{\textit{PAN19 zero-shot}} \\
R@20            & 0.097 & 0.179 \\
nDCG@20         & 0.029 & 0.064 \\
Triplet accuracy & \multicolumn{2}{c}{0.628} \\
\bottomrule
\end{tabular}
\caption{E5~\citep{wang_text_2024} zero-shot metrics on HALvest-Contrastive base-4 and PAN19, under dense and late-interaction scorers. Triplet accuracy is scorer-independent.}
\label{tab:e5-full}
\end{table}
 
We report E5 zero-shot metrics under both the dense scorer (cosine similarity of mean-pooled embeddings, the scoring mode E5 was trained for) and the late-interaction scorer (token-level $\mathrm{MaxSim}$, included as an ablation to verify that the conclusions of \secref{sec:exp-aa-ne-ir} are not an artifact of the chosen scorer).
 
The late-interaction scorer improves retrieval metrics uniformly over the dense scorer, but the overall trend remains the same: E5 retains a high triplet accuracy (0.789 on HALvest-C, 0.628 on PAN19) while struggling on retrieval. The gap between the two metric families is the motivating observation of \secref{sec:exp-aa-ne-ir}.

\section{Sequence length}
\label{app:seq_len}

\begin{figure}[ht]
  \centering
  \includegraphics[width=0.48\textwidth]{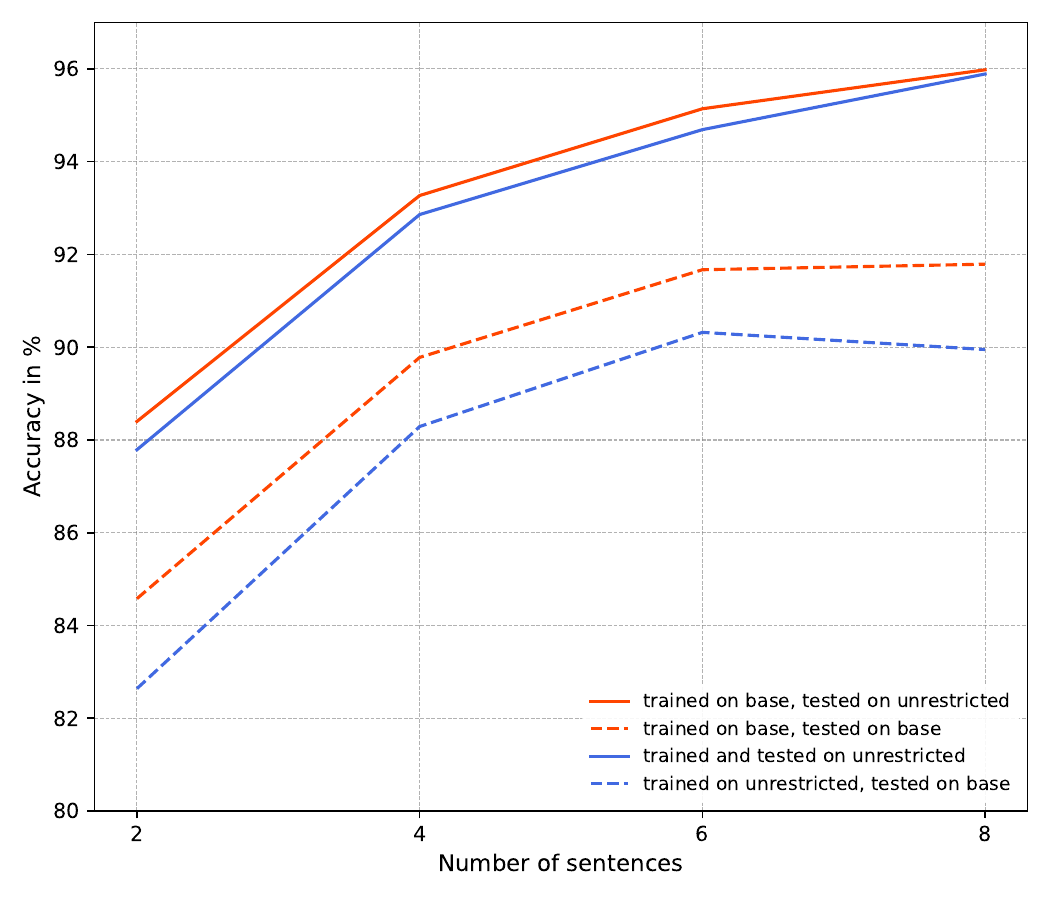}
  \caption{Two fine-tuned RoBERTa models, in blue and red, respectively trained on unrestricted and base data. Plain lines track performance on unrestricted test data.}
  \label{fig:model_performance}
\end{figure}

We observe that increasing sequence length from 2 to 8 sentences amplifies the lexical signal for both restricted and unrestricted triplet types. However, the effect is far more pronounced for unrestricted triplets. In contrast, the signal for restricted triplets, while also increasing, is substantially weaker and begins plateauing after 6 sentences. While more text provides more stylistic evidence, the amount of unique author-specific evidence in topic-decoupled scenarios appears to saturate quickly as shown in Figure~\ref{fig:model_performance}.

 \begin{figure*}[ht]
  \centering
  \includegraphics[width=\textwidth,height=0.15\textheight,keepaspectratio]{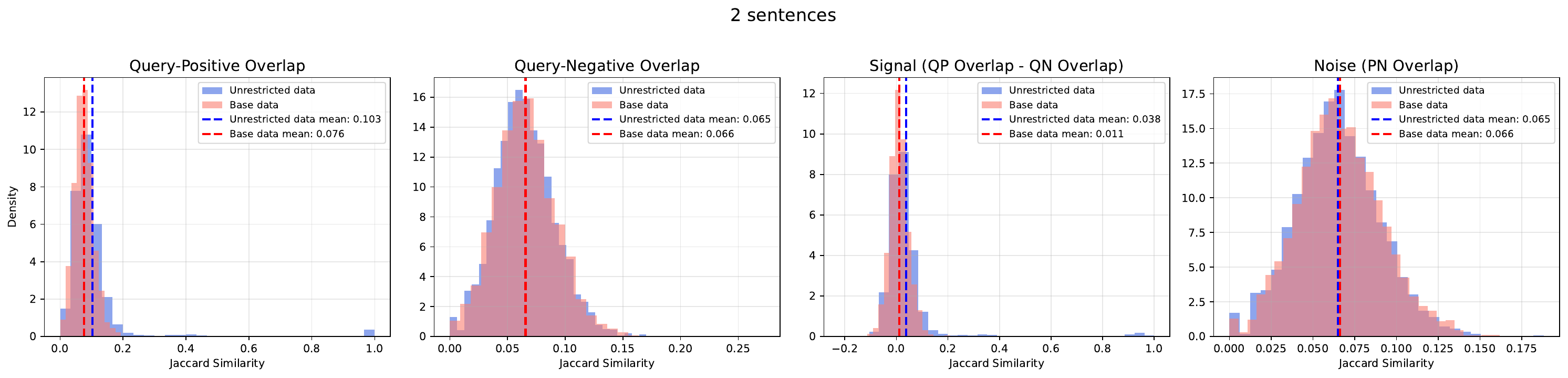}
  \includegraphics[width=\textwidth,height=0.15\textheight,keepaspectratio]{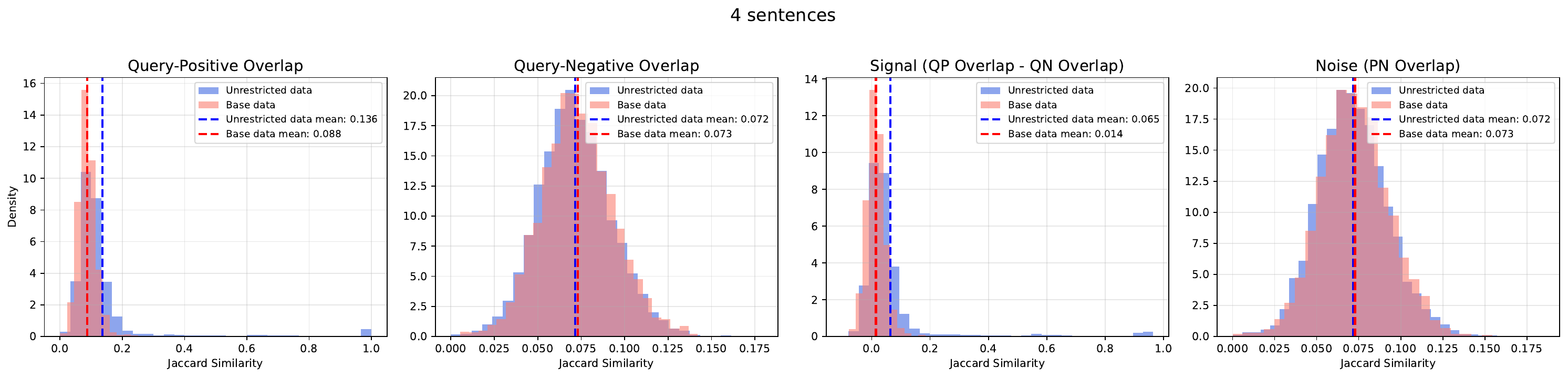}
  \includegraphics[width=\textwidth,height=0.15\textheight,keepaspectratio]{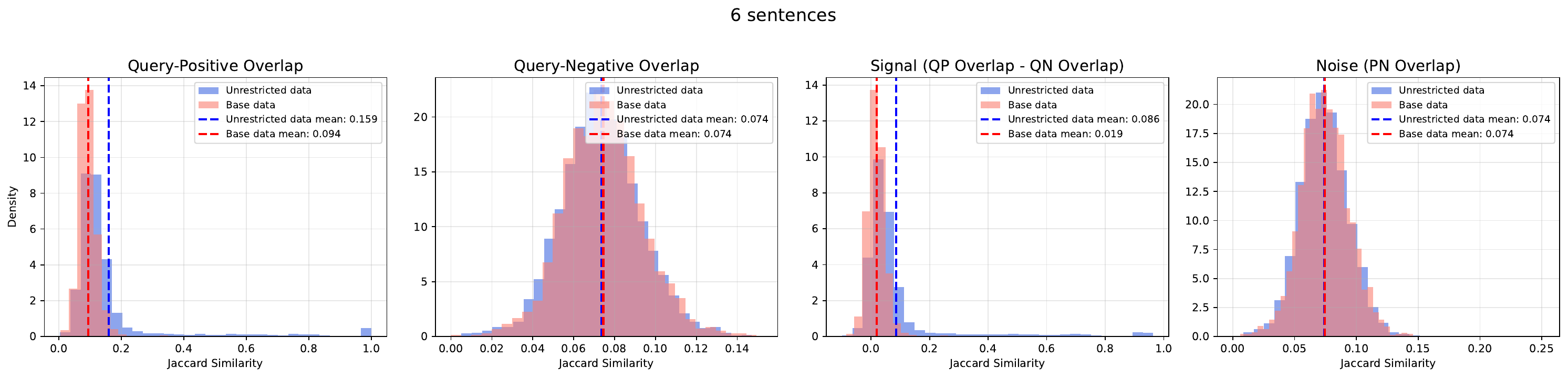}
  \includegraphics[width=\textwidth,height=0.15\textheight,keepaspectratio]{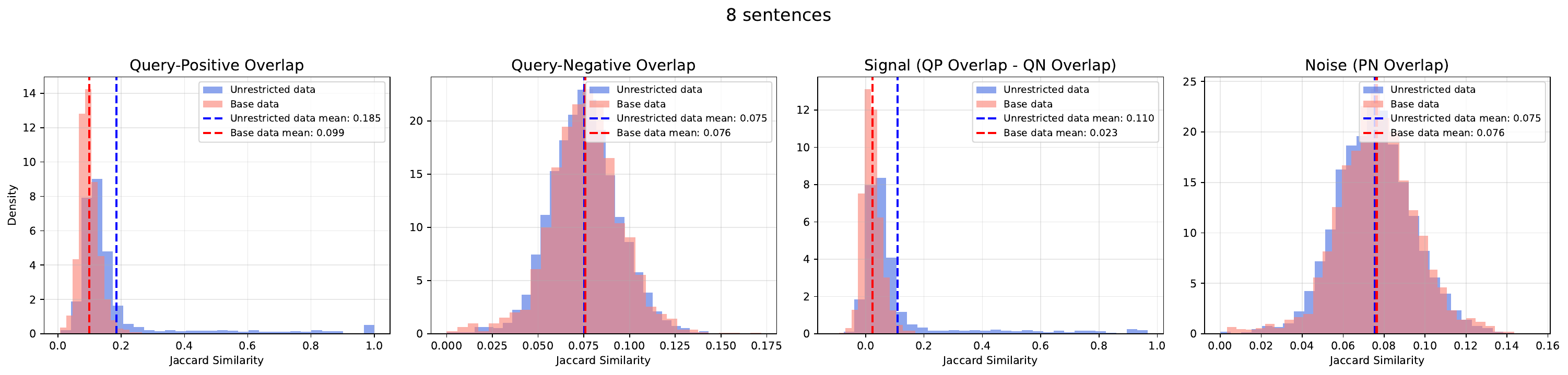}
  \caption{Signal and noise on every test split before scaling. We split 10,000 examples by word and compute Jaccard similarities. We define the signal as the difference between the query/positive and the query/negative overlaps whereas the noise is the Jaccard similarity between the positive and the negative of a triplet.}
  \label{fig:signal_noise}
\end{figure*}

We also analyzed the inherent lexical signal within contrastive triplets using Jaccard similarity before scaling HALvest-Contrastive. Recall that we tested three different configurations before settling on our current version of HALvest-Contrastive. Queries denotes what we call anchors in the main paper. As shown in Figure~\ref{fig:signal_noise}, the lexical signal is substantially higher in the unrestricted triplets, where the query and positive can come from the same paper. For the 8-sentence split, the mean signal is 0.11 for unrestricted triplets but only 0.022 for the base configuration. This confirms that base HALvest-Contrastive successfully minimizes keyword overlap, forcing any successful model to rely primarily on signals beyond shared tokens.


\section{Computational cost of late interaction and PLI}
\label{app:flops}
 
We derive the scoring cost of $\mathrm{MaxSim}$ for late interaction and for PLI with patch size~$n$, then instantiate the expressions under our training configuration.
 
\subsection{MaxSim cost per document pair}
 
Let $\mathbf{Q} \in \mathbb{R}^{S_q \times d}$ and $\mathbf{D} \in \mathbb{R}^{S_d \times d}$ be the token-embedding matrices of a query (anchor) and a document (candidate), where $S_q, S_d$ are the respective valid token counts and $d$ is the embedding dimension. Assuming $\ell_2$-normalised rows, $\mathrm{MaxSim}$ is computed as
 
\begin{equation}
\mathrm{MaxSim}(\mathbf{Q}, \mathbf{D}) = \sum_{i=1}^{S_q} \max_{j \in [S_d]} \; \mathbf{q}_i^\top \mathbf{d}_j \,.
\label{eq:maxsim-flops}
\end{equation}
 
The computation decomposes into three stages:
\begin{enumerate}
    \item \textbf{Pairwise dot products.} The full similarity matrix $\mathbf{Q}\mathbf{D}^\top \in \mathbb{R}^{S_q \times S_d}$ requires $S_q \cdot S_d$ inner products of dimension~$d$, each costing $2d - 1$ FLOPs ($d$ multiplications and $d{-}1$ additions). The total is $S_q S_d (2d - 1)$.
    \item \textbf{Row-wise maxima.} Extracting $\max_j$ for each of the $S_q$ rows requires $S_q(S_d - 1)$ comparisons.
    \item \textbf{Summation.} Summing the $S_q$ maxima costs $S_q - 1$ additions.
\end{enumerate}
 
Since $d \gg 1$, the pairwise dot-product term dominates and we write the cost as
\begin{equation}
C_{\mathrm{MaxSim}}(S_q, S_d, d) \;\approx\; 2\,S_q\,S_d\,d \,.
\label{eq:cost-maxsim}
\end{equation}
 
For the symmetric case $S_q = S_d = S$ common in our setup, this simplifies to $2S^2 d$.
 
\subsection{PLI: patching overhead and reduced scoring}
\label{app:flops-pli}
 
With a fixed patch size $n$, each sequence of $S$ tokens is compressed into $\mathcal{P} = \lceil S / n \rceil$ patch vectors by mean-pooling contiguous groups of $n$ token embeddings.
 
\paragraph{Patching cost.} For each of $\mathcal{P}$ patches, averaging $n$ vectors of dimension $d$ requires $(n{-}1) \cdot d$ additions and $d$ scalar divisions, totalling approximately $n \cdot d$ FLOPs per patch and $\mathcal{P} \cdot n \cdot d = S \cdot d$ FLOPs per sequence. For two sequences, the patching cost is $2Sd$.
 
\paragraph{Reduced MaxSim cost.} After patching, the similarity matrix is $\mathbf{Q}_{\mathrm{patch}} \mathbf{D}_{\mathrm{patch}}^\top \in \mathbb{R}^{P_q \times P_d}$, and the scoring cost becomes
 
\begin{equation}
C_{\mathrm{MaxSim}}^{\mathrm{PLI}}(S, d, n) \;\approx\; 2 \left\lceil \frac{S}{n} \right\rceil^2 d \;\approx\; \frac{2S^2 d}{n^2} \,.
\label{eq:cost-pli}
\end{equation}
 
The total PLI cost per pair is therefore $C_{\mathrm{MaxSim}}^{\mathrm{PLI}} + 2Sd$. Since $2Sd \ll 2S^2 d / n^2$ for all practical $S$ and $n \leq S$, the patching overhead is negligible and the scoring cost reduction factor is $n^2$.
 
\subsection{Scaling to a full training step}
\label{app:flops-training}
 
Under InfoNCE with full-gather, each anchor scores against every other document in the batch. With a batch size of $B = 256$, the positive is one of the $B$ candidates and the remaining $B - 1$ serve as negatives. Every document in the batch acts as an anchor in turn, so the total number of $\mathrm{MaxSim}$ evaluations per training step is $B^2$. Since each candidate's embedding matrix is computed once by the encoder and reused across all anchors, the $n^2$ speedup per-pair carries through unchanged: every one of the $B^2$ pairs gets cheaper by a factor of $n^2$, so the aggregate scoring cost drops by the same factor.
 
\begin{table}[ht]
\centering
\small
\begin{tabular}{lrrr}
\toprule
\textbf{Model} & \textbf{Patches/seq} & \textbf{FLOPs/pair} & \textbf{Speedup} \\
\midrule
LI (token)     & 512 & $4.03 \times 10^{8}$ & $1\times$ \\
PLI $n{=}2$    & 256 & $1.01 \times 10^{8}$ & $4\times$ \\
PLI $n{=}3$    & 171 & $4.49 \times 10^{7}$ & $\sim\!9\times$ \\
PLI $n{=}4$    & 128 & $2.52 \times 10^{7}$ & $16\times$ \\
PLI $n{=}5$    & 103 & $1.63 \times 10^{7}$ & $\sim\!25\times$ \\
\bottomrule
\end{tabular}
\caption{MaxSim scoring cost per document pair at $S = 512$, $d = 768$. The speedup column reports the ratio $C_{\mathrm{MaxSim}}^{\mathrm{LI}} / C_{\mathrm{MaxSim}}^{\mathrm{PLI}}$. The ``$\sim$'' prefix reflects the ceiling effect in $\lceil S/n \rceil$.}
\label{tab:flops}
\end{table}

Table~\ref{tab:flops} instantiates the cost per-pair for our primary configuration ($d = 768$) at the maximum sequence length $S = 512$.

\paragraph{Storage.} Each document requires $S \cdot d$ float values under LI and $\lceil S/n \rceil \cdot d$ under PLI, yielding a storage reduction factor of~$n$ (because storage scales with the number of vectors, not with pairwise interactions). At $S = 512$ and $d = 768$ in \texttt{float16}, a single document occupies 768\,KB under LI, 384\,KB under PLI $n{=}2$, and 154\,KB under PLI $n{=}5$.
 
\paragraph{Mixed sequence lengths.} Training uses a mix of span lengths (Table~\ref{tab:sqrt-law}), with mean token counts ranging from $\bar{S} \approx 70$ (base-2) to $\bar{S} \approx 328$ (base-10). Since $\mathrm{MaxSim}$ cost is quadratic in $S$, longer spans dominate the compute. The $n^2$ reduction factor applies uniformly regardless of $S$, so the relative advantage of PLI over LI is independent of the span-length distribution.

\end{document}